\documentclass[aps,prb,groupedaddress,onecolumn]{revtex4-1}
\pdfoutput=1
\usepackage{amssymb}
\usepackage{amsfonts}
\usepackage{amsmath}
\usepackage[pdftex]{graphicx}
\usepackage{dcolumn}
\usepackage{bm}
\usepackage{textcomp}
\usepackage{epstopdf}
\usepackage[colorlinks]{hyperref}
\hypersetup{
	colorlinks=true,
	linkcolor=blue,
	citecolor=blue,
	filecolor=blue,
	urlcolor=blue,
	pdfstartview={FitH},
}

\begin{document}

\title{Photoinduced inverse spin Hall effect in Pt/Ge(001) at room temperature}
\author{F. Bottegoni}
\email{federico.bottegoni@mail.polimi.it}
\author{A. Ferrari}
\author{S. Cecchi}
\author{M. Finazzi}
\author{F. Ciccacci}
\author{G. Isella}
\affiliation{LNESS-Dipartimento di Fisica, Politecnico di Milano, Piazza Leonardo da Vinci 32, 20133 Milano, Italy}
\date{\today }

\begin{abstract}
We performed photoinduced inverse spin Hall effect (ISHE) measurements on a Pt/Ge(001) junction at room temperature. The spin-oriented electrons, photogenerated at the direct gap of Ge using circularly polarized light, provide a net spin current which yields an electromotive field $\mathbf{E}_{\mathrm{ISHE}}$ in the Pt layer. Such a signal is clearly detected at room temperature despite the strong $\Gamma$ to $L$ scattering which electrons undergo in the Ge conduction band. The ISHE signal dependence on the exciting photon energy is in good agreement with the electron spin polarization expected for optical orientation at the direct gap of Ge.
\end{abstract}

\maketitle











The injection and detection of spin currents in solids represent the core of spintronics.\cite{Zutic2004} In direct gap III-V semiconductors a population of spin-oriented electrons can be generated at the $\Gamma$ point of the Brillouin zone upon optical excitation with circularly polarized light, a phenomenon known as \textit{optical orientation}.\cite{Lampel1968a} In bulk semiconductors the electron spin polarization, defined as $\mathbf{P}=\left(n_{\uparrow}-n_{\downarrow}\right)/
\left(n_{\uparrow}+n_{\downarrow}\right)\mathbf{u_{k}}$, with $n_{\uparrow}$ ($n_{\downarrow}$)
representing the densities of electrons with spin parallel (anti-parallel) to the quantization axis given by the direction of the incident light (parallel to the \textbf{u}$\mathbf{_{k}}$ unit vector), can reach $P=50$\% for resonant excitation at the $\Gamma$ point. It has been demonstrated that a significant optical spin orientation can also be achieved for the excitation at the direct bandgap of germanium\cite{Allenspach1983,Loren2009,Bottegoni2011} (E$_{g}^\Gamma=0.8$~eV at room temperature),
whose band structure at $\Gamma$ closely resembles that of III-V semiconductors.
Electrons photoexcited close to $\Gamma$ undergo a fast thermalization to the $L$ valleys within a time scale of about 300 fs, so that the electron transport in Ge generally occurs at the $L$ point.\cite{Zhou1994} However such $\Gamma$ to $L$ scattering partially preserves the initial electron spin, as observed in low-temperature photoluminescence  measurements\cite{Pezzoli2012} and at room temperature in spin-photodiodes exploiting magnetic tunnel junctions.\cite{Rinaldi2012} Recently, spin accumulation by spin-pumping has also been achieved in the Ge conduction band.\cite{Jain2012a} Moreover, relatively long electron spin-lifetimes have been predicted\cite{tang2012,Li2012} and experimentally measured \cite{Guite2011,Hautmann2011,Guite2012} in Ge. These findings, together with the "quasi direct" bandstructure of Ge and its compatibility with Si technology, make Ge a promising candidate for the implementation of devices where spins are optically injected and electrically detected, thus bridging the fields of spintronics and Si-photonics.

\begin{figure}[hb]
\centering
\includegraphics [width=\textwidth]{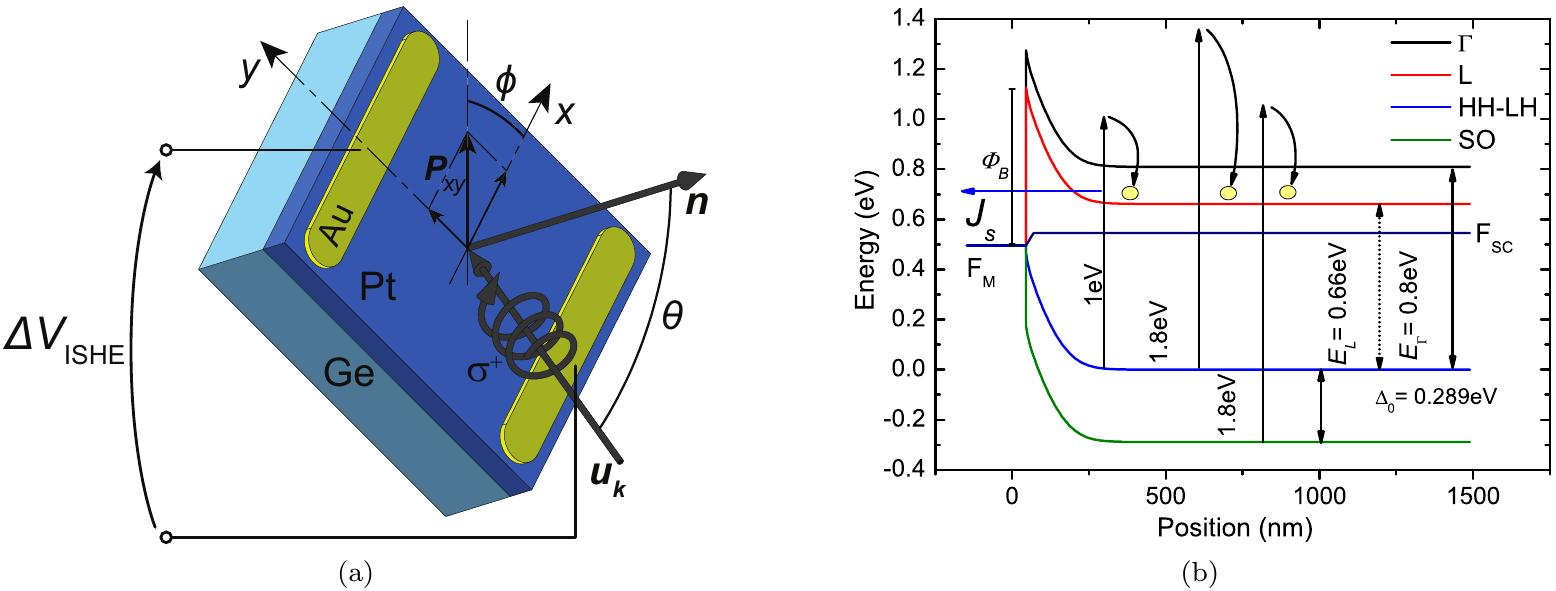}
\caption{\label{fig:1}(Color online) (a) Sketch of the Pt/Ge(001) sample, showing the relevant angles/directions in our experimental setup. The unit vector $\mathbf{u_{k}}$ indicates the direction of the incoming light.
(b) Schematic view of the bandstructure of the Pt/Ge interface. Spin-polarized electrons are excited from heavy-hole (HH), light-hole (LH) and split-off (SO) states to the Ge conduction band and, after relaxing to the $L$ minima, diffuse towards the Pt/Ge(001) Schottky barrier ($\Phi_{b}=0.63$~eV). A pure spin current $J_{s}$ is injected in the Pt layer.}
\end{figure}

In this paper we demonstrate that circularly polarized light can generate a net spin current in bulk Ge, which can be electrically detected by measuring (without any externally applied magnetic field) the inverse spin Hall effect (ISHE) induced in a Pt thin film deposited on a n-type Ge(001) crystal. The dependence of the ISHE signal on the incoming photon energy has also been investigated in the energy range $h\nu =1.0-1.8$~eV and interpreted within a model accounting for the initial polarization of optically oriented spins.

The experimental geometry is presented in Fig. \ref{fig:1}(a). The sample ($5$ $\times$ $5$ mm$^2$) consists of a 4 nm-thick Pt layer deposited by e-beam evaporation on a 450 $\mu$m-thick As-doped Ge substrate ($N_{D}= 1.7 \times 10^{16}$~cm$^{-3}$). The ISHE signal is measured as an electromotive force between two 50-nm-thick Au electrodes evaporated at the edges of the Pt layer. The experimental set-up is similar to the one used by Ando \textit{et al.}\cite{Ando2010} for the study of the photogenerated ISHE in a Pt/GaAs sample.

The sample was illuminated with a collimated monochromatic beam (spot size of about 4 mm) from a Ti-sapphire tunable laser, which provides photons in the spectral range from $1.2$ to $1.8$~eV.
A CW InAs-quantum dot laser was used for measurements at $1$~eV excitation energy. A photo-elastic modulator (PEM) operating at 50 kHz was exploited to modulate the light circular polarization. The differential voltage signal $\Delta V$ between the two Au electrodes (see Fig. \ref{fig:1}(a)) was detected by a lock-in amplifier.
The sample was mounted on a multi-axial stage which allowed the rotation of the sample around the polar angle $\theta$ and the azimuthal angle $\phi$ defined in Fig. \ref{fig:1}(a).

The ISHE provides the conversion of a spin current into an electromotive force through spin orbit interaction between the electrons diffusing in the Pt layer and the Pt nuclei.\cite{Kimura2007} In our case, the optical orientation process creates polarized electron-hole couples which, by ambipolar diffusion, reach the Pt/Ge(001) interface carrying a spin current but no \textit{charge} current (see Fig. \ref{fig:1}(b)).
In Ge the spin relaxation time for holes is $\tau_{sh}\approx 0.7$~ps at room temperature,\cite{Loren2011} much lower than the electron spin lifetime $\tau_{se}\approx 1$~ns,\cite{Guite2011,Hautmann2011,Guite2012} so that most of the spin current injected in the Pt layer is carried by electrons. A spin current $\mathbf{J}_{s}$ carrying a spin polarization $\mathbf{P}$ generates an electromotive field $\mathbf{E}_{\mathrm{ISHE}}$ given by:
\begin{equation}
\mathbf{E}_{\mathrm{ISHE}}=\frac{\gamma}{\sigma_{c}} \mathbf{J}_{s}\times\mathbf{P}
\label{eq:ISHE}
\end{equation}
\noindent where $\gamma=0.08$ and $\sigma_{c} = 1.1 \times 10^{6}$~($\Omega$ m)$^{-1}$ represent the spin-Hall angle and the electrical conductivity of Pt, respectively.\cite{Saitoh2006} Since $\mathbf{J}_{s}$ is perpendicular to the Pt/Ge interface and only the component of $\mathbf{E}_{\mathrm{ISHE}}$ along the direction connecting the two Au electrodes (\textit{y}-axis) can be detected, it is essential to maximize the $P_{x}$ component of the interfacial plane projection $\mathbf{P}_{xy}$ of the photoinduced spin polarization $\mathbf{P}$ (see Fig. \ref{fig:1}(a)) by illuminating the sample at grazing incidence.

\begin{figure}
\centering
\includegraphics[width=0.25\textwidth]{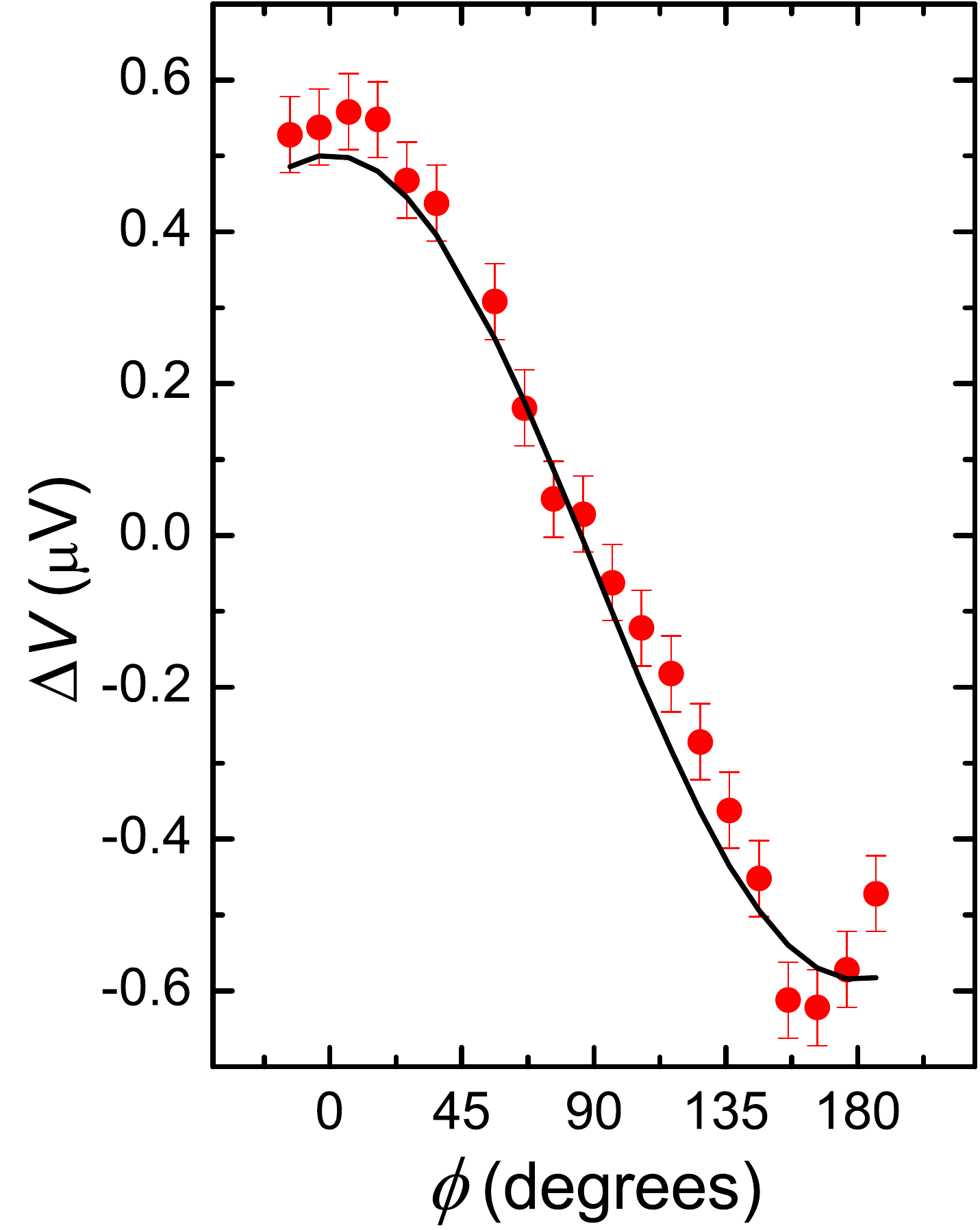}
\caption{\label{fig:sin} (Color online) ISHE electromotive force $\Delta V$ (red dots) as function of the angle $\phi$, where $\phi$ is the angle between $\mathbf{P}_{xy}$ (polarization vector $\mathbf{P}$ projection in the \textit{xy}-plane) and the \textit{x}-axis.  Data taken at $T = 300$~K. The experimental data are in good agreement with the expected $\cos\phi$ dependence (black curve).}
\end{figure}

Due to the relatively high refractive index of Ge ($n_{\mathrm{Ge}}\approx 5$), the angle of the light wave-vector inside the Ge crystal $\theta_{\mathrm{Ge}}$ is much smaller than the angle of incidence on the Pt layer $\theta$, resulting in a relatively small projection $\mathbf{P}_{xy}$ of $\mathbf{P}$. Moreover, due to Fresnel laws, circularly polarized light impinging on the Pt surface becomes elliptically polarized inside the Ge layer, thus reducing the spin polarization of the optically oriented spins. Neverthless, the ISHE signal can be shown to be proportional to the degree of circular polarization of the incoming beam $R_{c}$.\cite{Ando2010,Khamari2011}
Fig. \ref{fig:sin} shows the ISHE voltage as a function of $\phi$ for a fixed $\theta = 65^{\circ}$,  $h\nu=1.77$~eV and an excitation power $w_{exc} = 230$~mW. Measurements have been performed at room temperature with a noise voltage lower than 100 nV. $\Delta V$ follows the $\cos\phi$ dependence expected from Eq. \ref{eq:ISHE} with the experimental set-up depicted in Fig. \ref{fig:1}(a).

An intensity modulation of the exciting light would give a voltage signal, which should be detectable also in the second harmonic output of the lock-in amplifier, not related to the spin orientation but associated to the Dember effect \cite{Pankove}. Such a signal has been found to be negligibly small confirming the ISHE origin of $\Delta V$.

\begin{figure}
\centering
\includegraphics[width=.5\textwidth]{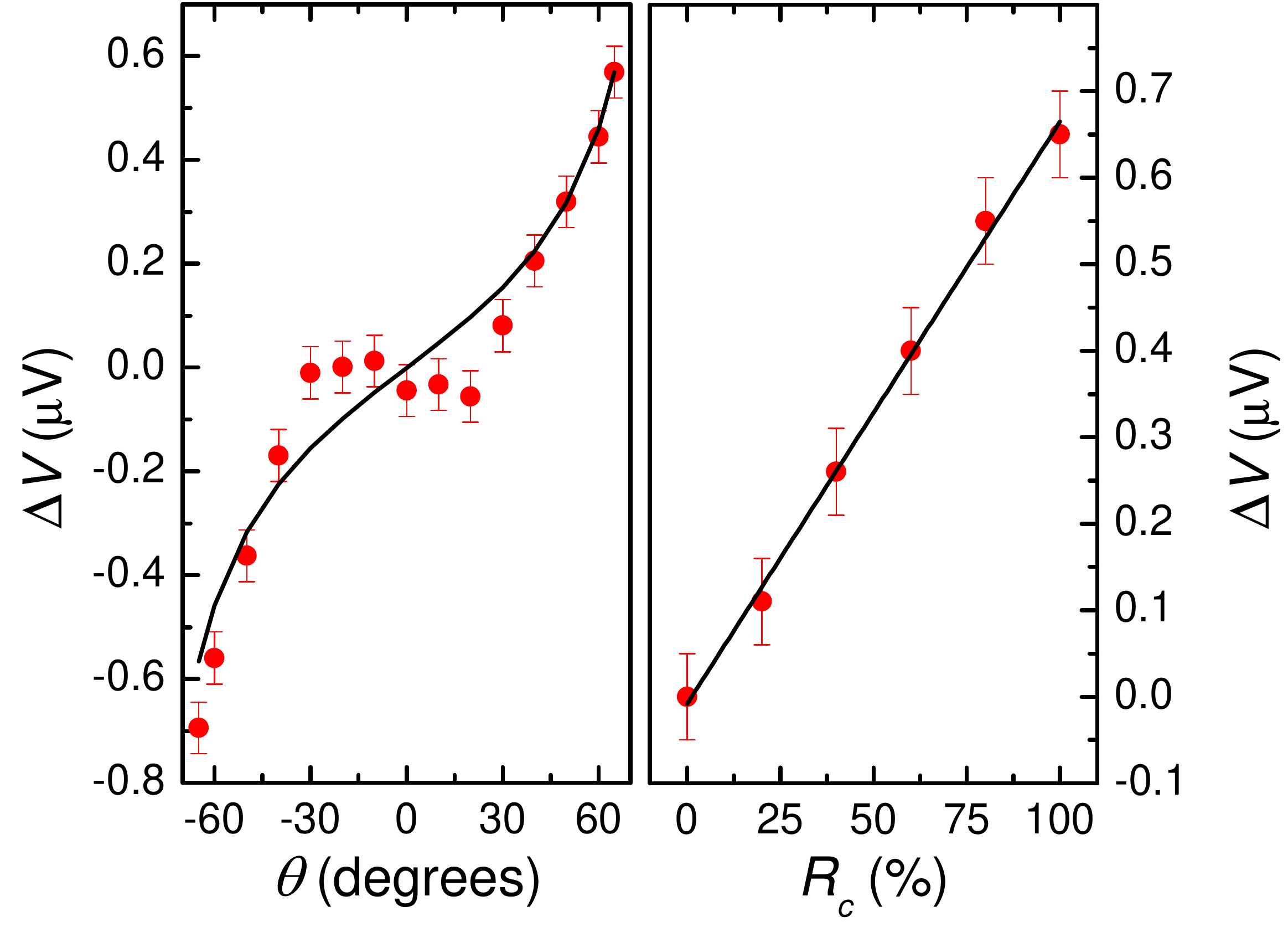}
\caption{\label{fig:3} (Color online) (a) ISHE voltage signal $\Delta V$ as function of the incidence angle $\theta$ at $T = 300$~K. (b) Dependence of $\Delta V$ on the degree of circular polarization $R_{c}$ of the exciting beam.}
\end{figure}

In order to validate the spin-current origin of the detected signal, we have performed additional measurements at $\phi=0^{\circ}$ as a function of the incidence angle $\theta$ (Fig. \ref{fig:3}(a)) and of the degree of circular polarization $R_{c}$ (Fig. \ref{fig:3}(b)).\footnote{The differential ISHE signal is related to the temporal modulation of the incoming incident light through the relation $\Delta V\sim \sin\left[\alpha \cos\left(\omega t\right)\right]$, where $\alpha$ is the phase-shift of the PEM and $\omega$ is the modulation frequency. The $\Delta V$ amplitude deconvoluted by the lock-in amplifier is related to the $\Delta V_{\mathrm{ISHE}}$ value that is measured by flipping the light circular polarization through the following formula: $\Delta V=2J_{\mathrm{1}}\left(\alpha\right)\Delta V_{\mathrm{ISHE}}$, with $J_{1}$ being the 1$^{st}$ Bessel Function.}
A variation in $\theta$ affects both the intensity and the degree of circular polarization of light reaching the Ge crystal, according to the relation $\Delta V_{\mathrm{ISHE}} \propto \cos\left(\theta_{\mathrm{Ge}}\right) \tan\theta R_{c}$. Fig. \ref{fig:3}(a) clearly shows that our data nicely reproduce the expected behavior (black curve in Fig. \ref{fig:3}(a)) within the experimental accuracy, except for low incidence angles where the ISHE signal is almost absent.
Fig. \ref{fig:3}(b) shows the dependence of $\Delta V$ (for $\theta=65^{\circ}$ and $\phi=0^{\circ}$) on the degree of circular polarization of the exciting beam. $R_{c}$ is varied changing the phase shift induced by the PEM from $0$ ($R_{c}=0$) to $\lambda/4$ ($R_{c}=100$\%). $\Delta V$ linearly increases with $R_{c}$, which is also expected considering Fresnel laws \cite{Ando2010,Khamari2011} and selection rules for optical orientation at the Ge direct gap.

In the experimental conditions where the sample is illuminated with a spot diameter $d$ roughly equal to the distance between the two Au electrodes, $d\approx 4$~mm, $\Delta V$ can be expressed as $\Delta V \approx E_{\mathrm{ISHE}}~d $. It is therefore possible to provide an estimation of the spin current injected at the Pt/Ge interface. By exploiting the spin-diffusion equations, as reported in Ref. \citenum{Ando2010}, we get:
\begin{align}
J_{s} & = \frac{t}{d}\frac{\sigma_{c}}{\gamma}\left(\int_0^t \frac{\sinh[(t-z)/\lambda]}{\sinh(t/\lambda)}dz\right)^{-1}\Delta V \notag \\
& \approx7.06\times10^9\Delta V(\mathrm{A} \cdot \mathrm{m}^{-2}) \label{current}
\end{align}
where $t=4$~nm is the Pt film thickness, $\lambda=7$~nm is the spin diffusion length in Pt,\cite{Saitoh2006} and $\Delta V$ is measured in volts. For $\theta = 65^{\circ}$, $\phi=0^{\circ}$, $h\nu=1.77$~eV and $w_{exc} = 230$~mW, we measure $\Delta V=0.48~\mu$V, which provides a spin current density equal to $J_{s} = 3 \times 10^{3}$~A~m$^{-2}$. Such a value, when normalized to the flux of incoming photons, is comparable to the one estimated in Ref. \citenum{Ando2010} for a Pt/GaAs(001) junction.

\begin{figure}[bt]
\includegraphics[width=0.3\textwidth]{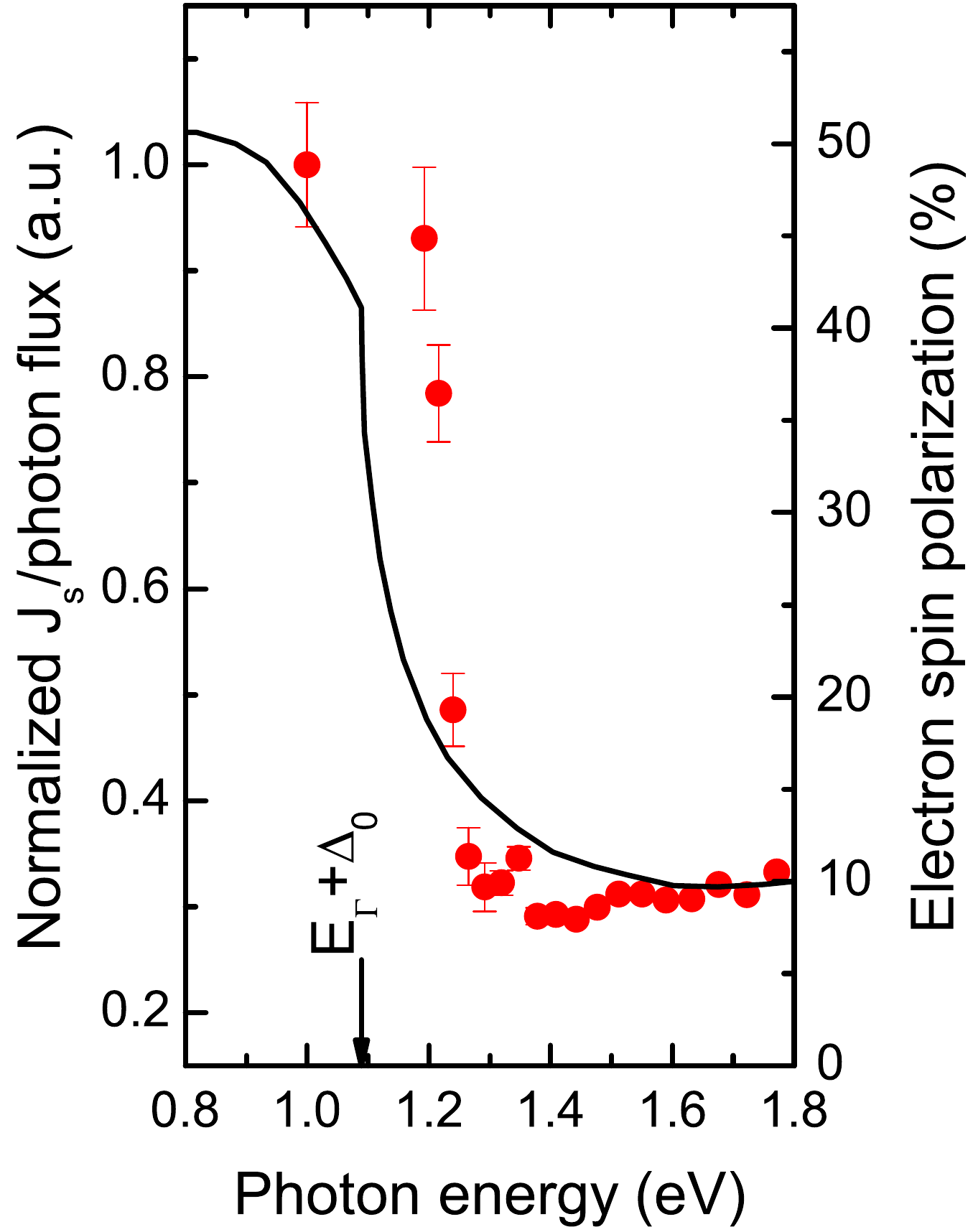}
\caption{\label{fig:disp} (Color online) Spin current $J_{s}(\theta=65^{\circ},\phi=0^{\circ})$ at the Pt/Ge(001) interface, normalized to the photon flux, as a function of photon energy (red dots, left axis), compared to the initial photo-induced electron spin polarization (black line, right axis) from Ref.~\citenum{Rioux2010}.}
\end{figure}

Fig. \ref{fig:disp} shows the spin current $J_{s}$, obtained from Eq. \ref{current} as a function of the incident photon energy $h\nu$ in the 1.0-1.8 eV range, normalized to the flux of the incoming photons.
In the absence of optical excitation, a dynamical equilibrium is established between the electron flux
between the semiconductor (SC) and the metal. Under illumination the barrier experienced by electrons in the SC is lowered by the open-circuit photo-voltage and the flux of electrons injected in the Pt layer increases. In our case such an electron flux is counterbalanced by the current transported by holes to ensure charge conservation. After excitation at the direct gap electrons relax to the $L$ minima. The lifetime of electrons in L- and $\Gamma$-states is $\tau_{L} \approx 1\mu$s and $\tau_{\Gamma}\approx 300$~fs, respectively. By applying a simple model taking into account the diffusion coefficient in Ge, one can readily estimate that a quasi-equilibrium regime is established,\cite{Grzybowski2011} where only a tiny fraction of carriers, of the order of $10^{-4}$ for an injected carrier density of $10^{19}$~cm$^{-3}$, occupies states at $\Gamma$. Carrier and spin diffusion is therefore dominated by transport through $L$-states.
The ambipolar diffusion coefficient in Ge is $D=63$~cm$^2$s$^{-1}$ and, assuming an electron spin lifetime of $\tau_{se}\approx 1$~ns, a spin diffusion length $L_{se}\approx 2.5~\mu$m can be calculated. The fraction of electrons, generated within a distance $L_{se}$ from the metal/SC interface, is not significantly varying within the photon energy range used in our experiments, being $87\%$ for $h\nu=1$~eV and $100\%$ for $h\nu=1.8$~eV. As a result, the normalized $J_{s}$ value is expected to be roughly proportional to the electron spin polarization $P$.
The degree of electron spin polarization $P_{0}$ right after excitation in the conduction band as calculated in Ref. \citenum{Rioux2010} is also shown in Fig. \ref{fig:disp} (continuous line, right-hand axis). The normalized $J_{s}$ value is seen to be roughly proportional to $P_{0}$, confirming that the ultrafast $\Gamma-L$ scattering partially preserves the electron spin, allowing its transport through the long-lived $L$ states.
It is worth noticing that the sharp decrease of the electron polarization expected for $h\nu = E_{\Gamma}+\Delta_{0}$ with $\Delta_0$ being the SO energy (black line in Fig. \ref{fig:disp}), is observed in the ISHE measurements at about 0.1 eV higher photon energies. Such a finding can be related to the different depth at which electrons promoted from the HH and SO bands are generated. As a first approximation, the absorption coefficient for direct gap transitions is proportional to $m_{r}^{(3/2)} \sqrt{h\nu-E_{t}}$, where $m_{r}$ is the valence/conduction band reduced mass for photoexcitation at $\Gamma$ and $E_{t}$ is the threshold energy of the transition. When exciting with right-circularly polarized photons having energy $h\nu=1.1$~eV, the absorption coefficient associated to the excitation of spin-down electrons from the SO band is roughly ten times smaller than the one related to the excitation of spin-up electrons from the HH band, due to the different values of $m_{r}$ and $E_{t}$ in the two transitions ($m_r=0.037$ and $E_{t}=E_{g}$ for HH to $\Gamma$, while $m_{r}=0.028$ and $E_{t}=E_{g}+\Delta_{0}$ for SO to $\Gamma$). Therefore, spin-down electrons are generated further away from the Pt/Ge(001) interface and give a minor contribution to the injected spin current.

In conclusion, we have shown that a detectable ISHE voltage signal from a Pt/Ge(001) junction can be obtained at room temperature. The attribution of the measured signal to ISHE is confirmed by its dependence on the incidence and azimuthal angles $\theta$ and $\phi$ and on the degree of polarization of the exciting light. The dispersion of the ISHE signal as a function of the exciting photon energy confirms that the optically-promoted electrons in the Ge conduction band can be transferred to the $L$ valleys still preserving, at least partially, their spin orientation. These results pave the way for the development of Ge-based devices exploiting photoinduced spin currents.

\begin{acknowledgments}
The EC is acknowledged for partially funding this work trough the GREEN Silicon project. The authors would like to acknowledge J. Frigerio and M. Bianchi for support during sample preparation and G. Bertuccio for fruitful discussions on optical excitation in Schottky barriers.
\end{acknowledgments}


\begin{thebibliography}{23}%
\makeatletter
\providecommand \@ifxundefined [1]{%
 \@ifx{#1\undefined}
}%
\providecommand \@ifnum [1]{%
 \ifnum #1\expandafter \@firstoftwo
 \else \expandafter \@secondoftwo
 \fi
}%
\providecommand \@ifx [1]{%
 \ifx #1\expandafter \@firstoftwo
 \else \expandafter \@secondoftwo
 \fi
}%
\providecommand \natexlab [1]{#1}%
\providecommand \enquote  [1]{``#1''}%
\providecommand \bibnamefont  [1]{#1}%
\providecommand \bibfnamefont [1]{#1}%
\providecommand \citenamefont [1]{#1}%
\providecommand \href@noop [0]{\@secondoftwo}%
\providecommand \href [0]{\begingroup \@sanitize@url \@href}%
\providecommand \@href[1]{\@@startlink{#1}\@@href}%
\providecommand \@@href[1]{\endgroup#1\@@endlink}%
\providecommand \@sanitize@url [0]{\catcode `\\12\catcode `\$12\catcode
  `\&12\catcode `\#12\catcode `\^12\catcode `\_12\catcode `\%12\relax}%
\providecommand \@@startlink[1]{}%
\providecommand \@@endlink[0]{}%
\providecommand \url  [0]{\begingroup\@sanitize@url \@url }%
\providecommand \@url [1]{\endgroup\@href {#1}{\urlprefix }}%
\providecommand \urlprefix  [0]{URL }%
\providecommand \Eprint [0]{\href }%
\providecommand \doibase [0]{http://dx.doi.org/}%
\providecommand \selectlanguage [0]{\@gobble}%
\providecommand \bibinfo  [0]{\@secondoftwo}%
\providecommand \bibfield  [0]{\@secondoftwo}%
\providecommand \translation [1]{[#1]}%
\providecommand \BibitemOpen [0]{}%
\providecommand \bibitemStop [0]{}%
\providecommand \bibitemNoStop [0]{.\EOS\space}%
\providecommand \EOS [0]{\spacefactor3000\relax}%
\providecommand \BibitemShut  [1]{\csname bibitem#1\endcsname}%
\let\auto@bib@innerbib\@empty
\bibitem [{\citenamefont {Zuti\'{c}}, \citenamefont {Fabian},\ and\
  \citenamefont {{Das Sarma}}(2004)}]{Zutic2004}%
  \BibitemOpen
  \bibfield  {author} {\bibinfo {author} {\bibfnamefont {I.}~\bibnamefont
  {Zuti\'{c}}}, \bibinfo {author} {\bibfnamefont {J.}~\bibnamefont {Fabian}}, \
  and\ \bibinfo {author} {\bibfnamefont {S.}~\bibnamefont {{Das Sarma}}},\
  }\href {\doibase 10.1103/RevModPhys.76.323} {\bibfield  {journal} {\bibinfo
  {journal} {Reviews of Modern Physics}\ }\textbf {\bibinfo {volume} {76}},\
  \bibinfo {pages} {323} (\bibinfo {year} {2004})},\ \Eprint
  {http://arxiv.org/abs/0405528} {arXiv:0405528 [cond-mat]} \BibitemShut
  {NoStop}%
\bibitem [{\citenamefont {Lampel}(1968)}]{Lampel1968a}%
  \BibitemOpen
  \bibfield  {author} {\bibinfo {author} {\bibfnamefont {G.}~\bibnamefont
  {Lampel}},\ }\href {\doibase 10.1103/PhysRevLett.20.491} {\bibfield
  {journal} {\bibinfo  {journal} {Physical Review Letters}\ }\textbf {\bibinfo
  {volume} {20}},\ \bibinfo {pages} {491} (\bibinfo {year} {1968})}\BibitemShut
  {NoStop}%
\bibitem [{\citenamefont {Allenspach}, \citenamefont {Meier},\ and\
  \citenamefont {Pescia}(1983)}]{Allenspach1983}%
  \BibitemOpen
  \bibfield  {author} {\bibinfo {author} {\bibfnamefont {R.}~\bibnamefont
  {Allenspach}}, \bibinfo {author} {\bibfnamefont {F.}~\bibnamefont {Meier}}, \
  and\ \bibinfo {author} {\bibfnamefont {D.}~\bibnamefont {Pescia}},\ }\href
  {\doibase 10.1103/PhysRevLett.51.2148} {\bibfield  {journal} {\bibinfo
  {journal} {Physical Review Letters}\ }\textbf {\bibinfo {volume} {51}},\
  \bibinfo {pages} {2148} (\bibinfo {year} {1983})}\BibitemShut {NoStop}%
\bibitem [{\citenamefont {Loren}\ \emph {et~al.}(2009)\citenamefont {Loren},
  \citenamefont {Ruzicka}, \citenamefont {Werake}, \citenamefont {Zhao},
  \citenamefont {van Driel},\ and\ \citenamefont {Smirl}}]{Loren2009}%
  \BibitemOpen
  \bibfield  {author} {\bibinfo {author} {\bibfnamefont {E.~J.}\ \bibnamefont
  {Loren}}, \bibinfo {author} {\bibfnamefont {B.~A.}\ \bibnamefont {Ruzicka}},
  \bibinfo {author} {\bibfnamefont {L.~K.}\ \bibnamefont {Werake}}, \bibinfo
  {author} {\bibfnamefont {H.}~\bibnamefont {Zhao}}, \bibinfo {author}
  {\bibfnamefont {H.~M.}\ \bibnamefont {van Driel}}, \ and\ \bibinfo {author}
  {\bibfnamefont {A.~L.}\ \bibnamefont {Smirl}},\ }\href {\doibase
  10.1063/1.3222869} {\bibfield  {journal} {\bibinfo  {journal} {Applied
  Physics Letters}\ }\textbf {\bibinfo {volume} {95}},\ \bibinfo {pages}
  {092107} (\bibinfo {year} {2009})}\BibitemShut {NoStop}%
\bibitem [{\citenamefont {Bottegoni}\ \emph {et~al.}(2011)\citenamefont
  {Bottegoni}, \citenamefont {Isella}, \citenamefont {Cecchi},\ and\
  \citenamefont {Ciccacci}}]{Bottegoni2011}%
  \BibitemOpen
  \bibfield  {author} {\bibinfo {author} {\bibfnamefont {F.}~\bibnamefont
  {Bottegoni}}, \bibinfo {author} {\bibfnamefont {G.}~\bibnamefont {Isella}},
  \bibinfo {author} {\bibfnamefont {S.}~\bibnamefont {Cecchi}}, \ and\ \bibinfo
  {author} {\bibfnamefont {F.}~\bibnamefont {Ciccacci}},\ }\href {\doibase
  10.1063/1.3599493} {\bibfield  {journal} {\bibinfo  {journal} {Applied
  Physics Letters}\ }\textbf {\bibinfo {volume} {98}},\ \bibinfo {pages}
  {242107} (\bibinfo {year} {2011})}\BibitemShut {NoStop}%
\bibitem [{\citenamefont {Zhou}, \citenamefont {van Driel},\ and\ \citenamefont
  {Mak}(1994)}]{Zhou1994}%
  \BibitemOpen
  \bibfield  {author} {\bibinfo {author} {\bibfnamefont {X.}~\bibnamefont
  {Zhou}}, \bibinfo {author} {\bibfnamefont {H.}~\bibnamefont {van Driel}}, \
  and\ \bibinfo {author} {\bibfnamefont {G.}~\bibnamefont {Mak}},\ }\href
  {\doibase 10.1103/PhysRevB.50.5226} {\bibfield  {journal} {\bibinfo
  {journal} {Physical Review B}\ }\textbf {\bibinfo {volume} {50}},\ \bibinfo
  {pages} {5226} (\bibinfo {year} {1994})}\BibitemShut {NoStop}%
\bibitem [{\citenamefont {Pezzoli}\ \emph {et~al.}(2012)\citenamefont
  {Pezzoli}, \citenamefont {Bottegoni}, \citenamefont {Trivedi}, \citenamefont
  {Ciccacci}, \citenamefont {Giorgioni}, \citenamefont {Li}, \citenamefont
  {Cecchi}, \citenamefont {Grilli}, \citenamefont {Song}, \citenamefont
  {Guzzi}, \citenamefont {Dery},\ and\ \citenamefont {Isella}}]{Pezzoli2012}%
  \BibitemOpen
  \bibfield  {author} {\bibinfo {author} {\bibfnamefont {F.}~\bibnamefont
  {Pezzoli}}, \bibinfo {author} {\bibfnamefont {F.}~\bibnamefont {Bottegoni}},
  \bibinfo {author} {\bibfnamefont {D.}~\bibnamefont {Trivedi}}, \bibinfo
  {author} {\bibfnamefont {F.}~\bibnamefont {Ciccacci}}, \bibinfo {author}
  {\bibfnamefont {A.}~\bibnamefont {Giorgioni}}, \bibinfo {author}
  {\bibfnamefont {P.}~\bibnamefont {Li}}, \bibinfo {author} {\bibfnamefont
  {S.}~\bibnamefont {Cecchi}}, \bibinfo {author} {\bibfnamefont
  {E.}~\bibnamefont {Grilli}}, \bibinfo {author} {\bibfnamefont
  {Y.}~\bibnamefont {Song}}, \bibinfo {author} {\bibfnamefont {M.}~\bibnamefont
  {Guzzi}}, \bibinfo {author} {\bibfnamefont {H.}~\bibnamefont {Dery}}, \ and\
  \bibinfo {author} {\bibfnamefont {G.}~\bibnamefont {Isella}},\ }\href
  {\doibase 10.1103/PhysRevLett.108.156603} {\bibfield  {journal} {\bibinfo
  {journal} {Physical Review Letters}\ }\textbf {\bibinfo {volume} {108}},\
  \bibinfo {pages} {1} (\bibinfo {year} {2012})}\BibitemShut {NoStop}%
\bibitem [{\citenamefont {Rinaldi}\ \emph {et~al.}(2012)\citenamefont
  {Rinaldi}, \citenamefont {Cantoni}, \citenamefont {Petti}, \citenamefont
  {Sottocorno}, \citenamefont {Leone}, \citenamefont {Caffrey}, \citenamefont
  {Sanvito},\ and\ \citenamefont {Bertacco}}]{Rinaldi2012}%
  \BibitemOpen
  \bibfield  {author} {\bibinfo {author} {\bibfnamefont {C.}~\bibnamefont
  {Rinaldi}}, \bibinfo {author} {\bibfnamefont {M.}~\bibnamefont {Cantoni}},
  \bibinfo {author} {\bibfnamefont {D.}~\bibnamefont {Petti}}, \bibinfo
  {author} {\bibfnamefont {A.}~\bibnamefont {Sottocorno}}, \bibinfo {author}
  {\bibfnamefont {M.}~\bibnamefont {Leone}}, \bibinfo {author} {\bibfnamefont
  {N.~M.}\ \bibnamefont {Caffrey}}, \bibinfo {author} {\bibfnamefont
  {S.}~\bibnamefont {Sanvito}}, \ and\ \bibinfo {author} {\bibfnamefont
  {R.}~\bibnamefont {Bertacco}},\ }\href {\doibase 10.1002/adma.201104256}
  {\bibfield  {journal} {\bibinfo  {journal} {Advanced Materials}\ ,\ \bibinfo
  {pages} {n/a}} (\bibinfo {year} {2012})}\BibitemShut {NoStop}%
\bibitem [{\citenamefont {Jain}\ \emph {et~al.}(2012)\citenamefont {Jain},
  \citenamefont {Rojas-Sanchez}, \citenamefont {Cubukcu}, \citenamefont
  {Peiro}, \citenamefont {{Le Breton}}, \citenamefont {Prestat}, \citenamefont
  {Vergnaud}, \citenamefont {Louahadj}, \citenamefont {Portemont},
  \citenamefont {Ducruet}, \citenamefont {Baltz}, \citenamefont {Barski},
  \citenamefont {Bayle-Guillemaud}, \citenamefont {Vila}, \citenamefont
  {Attan\'{e}}, \citenamefont {Augendre}, \citenamefont {Desfonds},
  \citenamefont {Gambarelli}, \citenamefont {Jaffr\`{e}s}, \citenamefont
  {George},\ and\ \citenamefont {Jamet}}]{Jain2012a}%
  \BibitemOpen
  \bibfield  {author} {\bibinfo {author} {\bibfnamefont {a.}~\bibnamefont
  {Jain}}, \bibinfo {author} {\bibfnamefont {J.-C.}\ \bibnamefont
  {Rojas-Sanchez}}, \bibinfo {author} {\bibfnamefont {M.}~\bibnamefont
  {Cubukcu}}, \bibinfo {author} {\bibfnamefont {J.}~\bibnamefont {Peiro}},
  \bibinfo {author} {\bibfnamefont {J.}~\bibnamefont {{Le Breton}}}, \bibinfo
  {author} {\bibfnamefont {E.}~\bibnamefont {Prestat}}, \bibinfo {author}
  {\bibfnamefont {C.}~\bibnamefont {Vergnaud}}, \bibinfo {author}
  {\bibfnamefont {L.}~\bibnamefont {Louahadj}}, \bibinfo {author}
  {\bibfnamefont {C.}~\bibnamefont {Portemont}}, \bibinfo {author}
  {\bibfnamefont {C.}~\bibnamefont {Ducruet}}, \bibinfo {author} {\bibfnamefont
  {V.}~\bibnamefont {Baltz}}, \bibinfo {author} {\bibfnamefont
  {A.}~\bibnamefont {Barski}}, \bibinfo {author} {\bibfnamefont
  {P.}~\bibnamefont {Bayle-Guillemaud}}, \bibinfo {author} {\bibfnamefont
  {L.}~\bibnamefont {Vila}}, \bibinfo {author} {\bibfnamefont {J.-P.}\
  \bibnamefont {Attan\'{e}}}, \bibinfo {author} {\bibfnamefont
  {E.}~\bibnamefont {Augendre}}, \bibinfo {author} {\bibfnamefont
  {G.}~\bibnamefont {Desfonds}}, \bibinfo {author} {\bibfnamefont
  {S.}~\bibnamefont {Gambarelli}}, \bibinfo {author} {\bibfnamefont
  {H.}~\bibnamefont {Jaffr\`{e}s}}, \bibinfo {author} {\bibfnamefont {J.-M.}\
  \bibnamefont {George}}, \ and\ \bibinfo {author} {\bibfnamefont
  {M.}~\bibnamefont {Jamet}},\ }\href {\doibase 10.1103/PhysRevLett.109.106603}
  {\bibfield  {journal} {\bibinfo  {journal} {Physical Review Letters}\
  }\textbf {\bibinfo {volume} {109}},\ \bibinfo {pages} {106603} (\bibinfo
  {year} {2012})}\BibitemShut {NoStop}%
\bibitem [{\citenamefont {Tang}, \citenamefont {Collins},\ and\ \citenamefont
  {Flatt\'{e}}(2012)}]{tang2012}%
  \BibitemOpen
  \bibfield  {author} {\bibinfo {author} {\bibfnamefont {J.-M.}\ \bibnamefont
  {Tang}}, \bibinfo {author} {\bibfnamefont {B.~T.}\ \bibnamefont {Collins}}, \
  and\ \bibinfo {author} {\bibfnamefont {M.~E.}\ \bibnamefont {Flatt\'{e}}},\
  }\href {\doibase 10.1103/PhysRevB.85.045202} {\bibfield  {journal} {\bibinfo
  {journal} {Physical Review B}\ }\textbf {\bibinfo {volume} {85}},\ \bibinfo
  {pages} {045202} (\bibinfo {year} {2012})}\BibitemShut {NoStop}%
\bibitem [{\citenamefont {Li}\ and\ \citenamefont {Appelbaum}(2012)}]{Li2012}%
  \BibitemOpen
  \bibfield  {author} {\bibinfo {author} {\bibfnamefont {J.}~\bibnamefont
  {Li}}\ and\ \bibinfo {author} {\bibfnamefont {I.}~\bibnamefont {Appelbaum}},\
  }\href {\doibase 10.1063/1.4704802} {\bibfield  {journal} {\bibinfo
  {journal} {Applied Physics Letters}\ }\textbf {\bibinfo {volume} {100}},\
  \bibinfo {pages} {162408} (\bibinfo {year} {2012})}\BibitemShut {NoStop}%
\bibitem [{\citenamefont {Guite}\ and\ \citenamefont
  {Venkataraman}(2011)}]{Guite2011}%
  \BibitemOpen
  \bibfield  {author} {\bibinfo {author} {\bibfnamefont {C.}~\bibnamefont
  {Guite}}\ and\ \bibinfo {author} {\bibfnamefont {V.}~\bibnamefont
  {Venkataraman}},\ }\href {\doibase 10.1103/PhysRevLett.107.166603} {\bibfield
   {journal} {\bibinfo  {journal} {Physical Review Letters}\ }\textbf {\bibinfo
  {volume} {107}},\ \bibinfo {pages} {4} (\bibinfo {year} {2011})}\BibitemShut
  {NoStop}%
\bibitem [{\citenamefont {Hautmann}, \citenamefont {Surrer},\ and\
  \citenamefont {Betz}(2011)}]{Hautmann2011}%
  \BibitemOpen
  \bibfield  {author} {\bibinfo {author} {\bibfnamefont {C.}~\bibnamefont
  {Hautmann}}, \bibinfo {author} {\bibfnamefont {B.}~\bibnamefont {Surrer}}, \
  and\ \bibinfo {author} {\bibfnamefont {M.}~\bibnamefont {Betz}},\ }\href
  {\doibase 10.1103/PhysRevB.83.161203} {\bibfield  {journal} {\bibinfo
  {journal} {Physical Review B}\ }\textbf {\bibinfo {volume} {83}},\ \bibinfo
  {pages} {1} (\bibinfo {year} {2011})}\BibitemShut {NoStop}%
\bibitem [{\citenamefont {Guite}\ and\ \citenamefont
  {Venkataraman}(2012)}]{Guite2012}%
  \BibitemOpen
  \bibfield  {author} {\bibinfo {author} {\bibfnamefont {C.}~\bibnamefont
  {Guite}}\ and\ \bibinfo {author} {\bibfnamefont {V.}~\bibnamefont
  {Venkataraman}},\ }\href {\doibase 10.1063/1.4772500} {\bibfield  {journal}
  {\bibinfo  {journal} {Applied Physics Letters}\ }\textbf {\bibinfo {volume}
  {101}},\ \bibinfo {pages} {252404} (\bibinfo {year} {2012})}\BibitemShut
  {NoStop}%
\bibitem [{\citenamefont {Ando}\ \emph {et~al.}(2010)\citenamefont {Ando},
  \citenamefont {Morikawa}, \citenamefont {Trypiniotis}, \citenamefont
  {Fujikawa}, \citenamefont {Barnes},\ and\ \citenamefont {Saitoh}}]{Ando2010}%
  \BibitemOpen
  \bibfield  {author} {\bibinfo {author} {\bibfnamefont {K.}~\bibnamefont
  {Ando}}, \bibinfo {author} {\bibfnamefont {M.}~\bibnamefont {Morikawa}},
  \bibinfo {author} {\bibfnamefont {T.}~\bibnamefont {Trypiniotis}}, \bibinfo
  {author} {\bibfnamefont {Y.}~\bibnamefont {Fujikawa}}, \bibinfo {author}
  {\bibfnamefont {C.~H.~W.}\ \bibnamefont {Barnes}}, \ and\ \bibinfo {author}
  {\bibfnamefont {E.}~\bibnamefont {Saitoh}},\ }\href {\doibase
  10.1063/1.3327809} {\bibfield  {journal} {\bibinfo  {journal} {Applied
  Physics Letters}\ }\textbf {\bibinfo {volume} {96}},\ \bibinfo {pages}
  {082502} (\bibinfo {year} {2010})}\BibitemShut {NoStop}%
\bibitem [{\citenamefont {Kimura}\ \emph {et~al.}(2007)\citenamefont {Kimura},
  \citenamefont {Otani}, \citenamefont {Sato}, \citenamefont {Takahashi},\ and\
  \citenamefont {Maekawa}}]{Kimura2007}%
  \BibitemOpen
  \bibfield  {author} {\bibinfo {author} {\bibfnamefont {T.}~\bibnamefont
  {Kimura}}, \bibinfo {author} {\bibfnamefont {Y.}~\bibnamefont {Otani}},
  \bibinfo {author} {\bibfnamefont {T.}~\bibnamefont {Sato}}, \bibinfo {author}
  {\bibfnamefont {S.}~\bibnamefont {Takahashi}}, \ and\ \bibinfo {author}
  {\bibfnamefont {S.}~\bibnamefont {Maekawa}},\ }\href {\doibase
  10.1103/PhysRevLett.98.156601} {\bibfield  {journal} {\bibinfo  {journal}
  {Physical Review Letters}\ }\textbf {\bibinfo {volume} {98}},\ \bibinfo
  {pages} {1} (\bibinfo {year} {2007})}\BibitemShut {NoStop}%
\bibitem [{\citenamefont {Loren}\ \emph {et~al.}(2011)\citenamefont {Loren},
  \citenamefont {Rioux}, \citenamefont {Lange}, \citenamefont {Sipe},
  \citenamefont {van Driel},\ and\ \citenamefont {Smirl}}]{Loren2011}%
  \BibitemOpen
  \bibfield  {author} {\bibinfo {author} {\bibfnamefont {E.}~\bibnamefont
  {Loren}}, \bibinfo {author} {\bibfnamefont {J.}~\bibnamefont {Rioux}},
  \bibinfo {author} {\bibfnamefont {C.}~\bibnamefont {Lange}}, \bibinfo
  {author} {\bibfnamefont {J.}~\bibnamefont {Sipe}}, \bibinfo {author}
  {\bibfnamefont {H.}~\bibnamefont {van Driel}}, \ and\ \bibinfo {author}
  {\bibfnamefont {A.}~\bibnamefont {Smirl}},\ }\href {\doibase
  10.1103/PhysRevB.84.214307} {\bibfield  {journal} {\bibinfo  {journal}
  {Physical Review B}\ }\textbf {\bibinfo {volume} {84}},\ \bibinfo {pages} {1}
  (\bibinfo {year} {2011})}\BibitemShut {NoStop}%
\bibitem [{\citenamefont {Saitoh}\ \emph {et~al.}(2006)\citenamefont {Saitoh},
  \citenamefont {Ueda}, \citenamefont {Miyajima},\ and\ \citenamefont
  {Tatara}}]{Saitoh2006}%
  \BibitemOpen
  \bibfield  {author} {\bibinfo {author} {\bibfnamefont {E.}~\bibnamefont
  {Saitoh}}, \bibinfo {author} {\bibfnamefont {M.}~\bibnamefont {Ueda}},
  \bibinfo {author} {\bibfnamefont {H.}~\bibnamefont {Miyajima}}, \ and\
  \bibinfo {author} {\bibfnamefont {G.}~\bibnamefont {Tatara}},\ }\href
  {\doibase 10.1063/1.2199473} {\bibfield  {journal} {\bibinfo  {journal}
  {Applied Physics Letters}\ }\textbf {\bibinfo {volume} {88}},\ \bibinfo
  {pages} {182509} (\bibinfo {year} {2006})}\BibitemShut {NoStop}%
\bibitem [{\citenamefont {Khamari}, \citenamefont {Dixit},\ and\ \citenamefont
  {Oak}(2011)}]{Khamari2011}%
  \BibitemOpen
  \bibfield  {author} {\bibinfo {author} {\bibfnamefont {S.~K.}\ \bibnamefont
  {Khamari}}, \bibinfo {author} {\bibfnamefont {V.~K.}\ \bibnamefont {Dixit}},
  \ and\ \bibinfo {author} {\bibfnamefont {S.~M.}\ \bibnamefont {Oak}},\ }\href
  {\doibase 10.1088/0022-3727/44/26/265104} {\bibfield  {journal} {\bibinfo
  {journal} {Journal of Physics D: Applied Physics}\ }\textbf {\bibinfo
  {volume} {44}},\ \bibinfo {pages} {265104} (\bibinfo {year}
  {2011})}\BibitemShut {NoStop}%
\bibitem [{\citenamefont {Pankove}(1975)}]{Pankove}%
  \BibitemOpen
  \bibfield  {author} {\bibinfo {author} {\bibfnamefont {J.~I.}\ \bibnamefont
  {Pankove}},\ }\href@noop {} {\emph {\bibinfo {title} {Optical processes in
  semiconductors}}}\ (\bibinfo  {publisher} {Dover},\ \bibinfo {address} {New
  York, NY},\ \bibinfo {year} {1975})\BibitemShut {NoStop}%
\bibitem [{Note1()}]{Note1}%
  \BibitemOpen
  \bibinfo {note} {The differential ISHE signal is related to the temporal
  modulation of the incoming incident light through the relation $\Delta V\sim
  \protect \qopname \relax o{sin}\left [\alpha \protect \qopname \relax
  o{cos}\left (\omega t\right )\right ]$, where $\alpha $ is the phase-shift of
  the PEM and $\omega $ is the modulation frequency. The $\Delta V$ amplitude
  deconvoluted by the lock-in amplifier is related to the $\Delta V_{\protect
  \mathrm {ISHE}}$ value that is measured by flipping the light circular
  polarization through the following formula: $\Delta V=2J_{\protect \mathrm
  {1}}\left (\alpha \right )\Delta V_{\protect \mathrm {ISHE}}$, with $J_{1}$
  being the 1$^{st}$ Bessel Function.}\BibitemShut {Stop}%
\bibitem [{\citenamefont {Rioux}\ and\ \citenamefont {Sipe}(2010)}]{Rioux2010}%
  \BibitemOpen
  \bibfield  {author} {\bibinfo {author} {\bibfnamefont {J.}~\bibnamefont
  {Rioux}}\ and\ \bibinfo {author} {\bibfnamefont {J.~E.}\ \bibnamefont
  {Sipe}},\ }\href {\doibase 10.1103/PhysRevB.81.155215} {\bibfield  {journal}
  {\bibinfo  {journal} {Physical Review B}\ }\textbf {\bibinfo {volume} {81}},\
  \bibinfo {pages} {27} (\bibinfo {year} {2010})}\BibitemShut {NoStop}%
\bibitem [{\citenamefont {Grzybowski}\ \emph {et~al.}(2011)\citenamefont
  {Grzybowski}, \citenamefont {Roucka}, \citenamefont {Mathews}, \citenamefont
  {Jiang}, \citenamefont {Beeler}, \citenamefont {Kouvetakis},\ and\
  \citenamefont {Men\'{e}ndez}}]{Grzybowski2011}%
  \BibitemOpen
  \bibfield  {author} {\bibinfo {author} {\bibfnamefont {G.}~\bibnamefont
  {Grzybowski}}, \bibinfo {author} {\bibfnamefont {R.}~\bibnamefont {Roucka}},
  \bibinfo {author} {\bibfnamefont {J.}~\bibnamefont {Mathews}}, \bibinfo
  {author} {\bibfnamefont {L.}~\bibnamefont {Jiang}}, \bibinfo {author}
  {\bibfnamefont {R.}~\bibnamefont {Beeler}}, \bibinfo {author} {\bibfnamefont
  {J.}~\bibnamefont {Kouvetakis}}, \ and\ \bibinfo {author} {\bibfnamefont
  {J.}~\bibnamefont {Men\'{e}ndez}},\ }\href {\doibase
  10.1103/PhysRevB.84.205307} {\bibfield  {journal} {\bibinfo  {journal}
  {Physical Review B}\ }\textbf {\bibinfo {volume} {84}},\ \bibinfo {pages}
  {205307} (\bibinfo {year} {2011})}\BibitemShut {NoStop}%
\end{thebibliography}
\end{document}